\documentclass{article} 

\usepackage[utf8]{inputenc}
\usepackage{setspace}
\usepackage{algorithm}
\usepackage{algorithmic}
\usepackage{url}
\usepackage{ulem}
\usepackage[usenames,dvipsnames,svgnames,table]{xcolor}
\usepackage{hyperref}
\hypersetup{
	unicode,
	colorlinks,
	urlcolor=cyan, 
	linkcolor=blue, 
	citecolor=violet,
	pdftitle={libRoadRunner 2.0:},
	pdfsubject={SBML Simulator},
	pdfkeywords={SBML, Systems Biology},
	pdfproducer={LaTeX},
	pdfcreator={pdflatex}
}
\usepackage{cleveref}
\newcommand{\comment}[1]{}
\usepackage{graphicx, color}
\usepackage[margin=1in]{geometry} 

\usepackage{natbib}
\title{libRoadRunner 2.0: A High Performance SBML Simulation and Analysis Library}

\author{Ciaran Welsh$^{1}$ \and Jin Xu$^{2}$ \and
Lucian Smith$^{1}$ \and Matthias K\"{o}nig$^{2}$ \and Kiri Choi$^{3}$ \and Herbert M. Sauro$^{1}$}

\date{
	$^{1}$Department of Bioengineering, University of Washington, Seattle, WA 98195, USA,\\
	$^{2}$Department of Computational Systems Biochemistry,\\
  University Medicine Charit\'{e}  Berlin, 10117 Berlin, Germany,\\
  $^{3}$School of Computational Sciences, Korea Institute for Advanced Study, 02455 Seoul, Republic of Korea.}

\begin{document}
\maketitle

\begin{abstract}
{\textbf Motivation:} This paper presents libRoadRunner 2.0, an extensible, high-performance, cross-platform, open-source software library for the simulation and analysis of models expressed using Systems Biology Markup Language (\textit{SBML}).

\textbf{Results:} libRoadRunner is a self-contained library, able to run both as a component inside other tools via its C++ and C bindings, and interactively through its Python or Julia interface. libRoadRunner uses a custom Just-In-Time (\textit{JIT}) compiler built on the widely-used LLVM JIT compiler framework. It compiles SBML-specified models directly into native machine code for a large variety of processors, making it appropriate for solving extremely large models or repeated runs. libRoadRunner is flexible, supporting the bulk of the SBML specification (except for delay and nonlinear algebraic equations) and including several SBML extensions such as composition and distributions. It offers multiple deterministic and stochastic integrators, as well as tools for steady-state, sensitivity, stability analysis and structural analysis of the stoichiometric matrix.

\textbf{Availability:} libRoadRunner binary distributions are available for Mac OS X, Linux and Windows. The library is licensed under the Apache License Version 2.0. libRoadRunner is also available for ARM based computers such as the Raspberry Pi and can in principle be compiled on any system supported by LLVM-13. \url{http://sys-bio.github.io/roadrunner/index.html} provides online documentation, full build instructions, binaries and a git source repository.
\textbf{Contact:}\href{hsauro@uw.edu}{hsauro@uw.edu}
\end{abstract}

\section{Introduction}
Dynamic network models~\citep{SauroBookPathwayModeling} of metabolic, gene regulatory, protein signaling and electrophysiological models require the specification of components, interactions, compartments and kinetic parameters. The Systems Biology Markup Language (\textit{SBML})~\citep{Hucka:2003fs} has become the \textit{de facto} standard for declarative specification of these types of model (see SBML.org and refs.).

Popular tools for the development, simulation and analysis of models specified in SBML include COPASI~\citep{Hoops:2006ui}, Systems Biology Workbench (\textit{SBW})~\citep{Bergmann:2006ut}, The Systems Biology Simulation Core Algorithm (\textit{TSBSC})~\citep{Keller:2013cn}, Jarnac~\citep{sauro2000jarnac}, libSBMLSim~\citep{Takizawa:2013tj}, SOSLib~\citep{machne2006sbml}, iBioSim~\citep{iBioSim:2009}, PySCeS~\citep{Pysces:2005}, and VirtualCell~\citep{Moraru:2008iv}. Some of these applications are stand-alone packages designed for interactive use, with limited reusability as components in other applications. Very few are reusable libraries. Currently, none are fast enough to support emerging applications that require large-scale simulation of network dynamics. For example, multicell virtual-tissue simulations~\citep{Hester:2011tz} often require simultaneous simulation of tens of thousands of replicas of models residing in their cell objects and interacting between cells. In addition, optimization methods require generation of time-series for tens of thousands of replicas to explore the high-dimensional parameter spaces typical of biochemical networks~\citep{bouteiller2015maximizing}.

Previously we published libRoadRunner v1.0, a cross-platform, multi-language library for fast execution of SBML model simulations. We designed libRoadRunner to provide: 1) Efficient time-series generation and analysis of large or multiple SBML-based models; 2) A comprehensive and logical API; 3) Interactive simulations in the style of IPython and MATLAB; and 4) Extensibility. libRoadRunner achieves its performance capabilities by compiling SBML directly into machine code "on-the-fly` using LLVM as a "just-in-time` (JIT) compiler ~\citep{Lattner:2004vw}. The SBML model description is lexed and parsed into an abstract syntax tree (AST) using libSBML. From here libRoadRunner creates the necessary low level LLVM intermediate representation (IR) code for compiling the SBML. Once compiled, the SBML representation of the model has been converted into an in-memory dynamic library from which symbols representing model functions can be exported and loaded into other languages. libRoadRunner wraps this low level interface in a user friendly API in C++, which in turn provides the foundation for critical systems modelling tasks, such as model integration, steady state analysis and metabolic control analysis ~\citep{somogyi2015libroadrunner}.

libRoadRunner users usually fall into one of two categories: modellers or tool developers. Modellers use libRoadRunner tool directly in their research for modelling dynamic systems \citep{karagoz2021win} (more???) or developing new computational approaches such as detecting bistable switches \citep{reyes2022numerical} (more???). Tool developers on the other hand use libRoadRunner as a core SBML handling component in their modular software design \citep{choi2018tellurium}, runBiosimulations  \citep{shaikh2021runbiosimulations}, MASSPy \citep{haiman2021masspy}, SBMLUtils \citep{watanabe2018dynamic}, Compucell3D \citep{swat2012multi}, PhysiCell \citep{ghaffarizadeh2018physicell}, pyBioNetFit \citep{neumann2021implementation} and DIVIPAC \citep{nguyen2015dyvipac}.

In this work we present libRoadRunner version 2. We have improved performance both for single model simulations and multi-model simulations. We have expanded the range of available features to include additional steady state solvers, time series a sensitivity.

\section{Major Changes to 2.0}
\subsection{Performance Improvements}
In previous versions of libRoadRunner, loading many RoadRunner instances was slow because each model must JIT compile SBML to binary. We have addressed this problem in several ways: 1) by increasing the speed of compiling a single model; 2) by making it easy to compile many models simultaneously and 3) by providing a "direct` API for access to the model topology outside of modifying the SBML directly.

\subsubsection{LLJit: A New JIT Compiler}

To increase the speed of compiling SBML to machine code we have built a new JIT compiler called LLJit. LLJit uses LLVM version 13's "ORC JIT v2" API which provides an out of the box but modular and customizable tool for JIT compiling LLVM IR code to machine code. To be clear, was not necessary to modify the LLVM Intermediate Representation (IR) generation stage of the compilation process but we just use a new strategy for performing the compilation step. Our implementation of LLJit uses the standard object linking layer but a customized compile layer which automatically caches model object files in memory fast reloading. Switching to the LLJit compiler is shown in the listing below.

\begin{verbatim}
from roadrunner import RoadRunner, Config
Config.setValue(
    Config.LLVM_BACKEND, Config.LLJIT)
r = RoadRunner(sbmlFile)
\end{verbatim}

Python example of how to turn on the LLJit compiler. Variables: sbmlFile is absolute path to sbml file on disk.

%
\subsubsection{RoadRunnerMap: A parallel RoadRunner container}
As already stated, RoadRunner models are computationally expensive to compile. In addition to improving compile time for a single model, we have made it easy for users to make use of their multi-cored system for compiling multiple models in parallel. libRoadRunner uses a lightweight abstraction around the standard C++ 17 threading library called thread\_pool \citep{shoshany2021cpp} for queuing build jobs and then storing references to compiled RoadRunner models in a thread-safe hash map structure called RoadRunnerMap.
\begin{verbatim}
from roadrunner import RoadRunnerMap
rrm = RoadRunnerMap(listOfSBML, 3)
\end{verbatim}
Python example of loading a list of SBML models in parallel using three threads. Variables: listOfSBML is a list of full paths to SBML files on disc or strings in memory (or a mix thereof).

\begin{figure}[thp]
    \centering
    \includegraphics[width=0.45\textwidth]{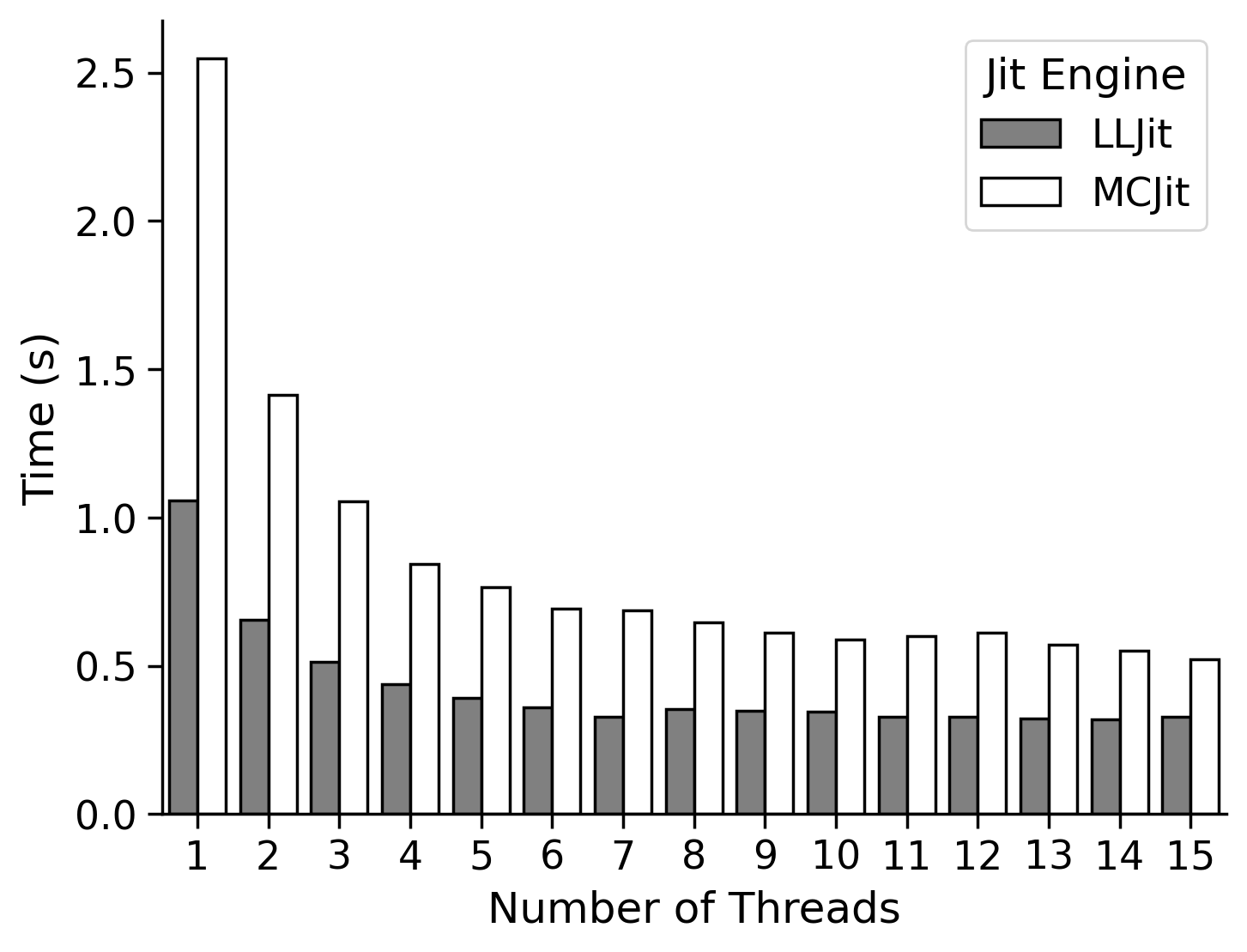}
    \caption{Time taken for {RoadRunnerMap} to load the first 100 models from the SBML test suite using the specified number of threads. Shown is the average of 10 replicates and errors are standard error of the mean (SEM).}
    \label{fig:CompileTime}
\end{figure}

%
To construct a RoadRunnerMap, pass a collection of SBML files or strings to the RoadRunnerMap constructor, along with an integer specifying the number of threads to use. To demonstrate the capabilities of RoadRunnerMap we compare the speed with which 100 models from the SBML test suite can be compiled using differing numbers of threads. As expected, increasing the number of threads decreases runtime but with diminishing returns \cref{fig:CompileTime}.
\subsubsection{Pickled (Serialized) RoadRunner}
Once loaded, users can save a model's state either to a binary string for in-memory storage or to disc for persistent storage. The result is a platform-specific binary snapshot of a RoadRunner object which can be reloaded with significant performance improvements compared to recompiling the model.
\begin{verbatim}
from roadrunner import RoadRunner
rr = RoadRunner(sbmlFile)
# save state to string
rr.saveState(fileName)
# load state
rrReloaded = RoadRunner()
rrReloaded.loadState(fileName)
\end{verbatim}

Example of saving a {RoadRunner} object's state to file and then loading it again. Variables: {sbmlFile} is a full path to a valid SBML file on disc; {fileName} is a full path to where to save the model state on disc.

%
A prerequisite for using {RoadRunner} with various parallel or multithreading toolboxes in Python is the ability to serialize an instance of a {RoadRunner} object using Python's standardized "pickle" protocol. We have built an adaptor between our in-house {RoadRunner} serialization strategy and Python's pickle protocol so that our users can now build their own parallel applications on top of libRoadRunner. We anticipate that this will be valuable to the systems biology community, particularly for problems involving repeated time series simulations such as optimization or stochastic simulations.  
\begin{verbatim}
from multiprocessing import Pool
from roadrunner import RoadRunner
from roadrunner import RoadRunner
def simulate_worker(r: RoadRunner):
    r.resetAll()
    return r.simulate(0, 10, 11)
r = RoadRunner(sbmlFile)
r.setIntegrator('gillespie')
p = Pool(processes=8)
results = p.map(
    simulate_worker, 
    [r for i in range(100000)])
\end{verbatim}

Example of using {RoadRunner} object with Python's multiprocessing library to simulate a model stochastically 100K times.
%
\subsection{Direct API}
In libRoadRunner version 1, any changes to the sbml requires the modification, re-parsing and then re-compiling the SBML. Since these are expensive operations, we have implemented an API for interacting directly with the model topology. The so called "direct` API allows users to add and remove SBML components such as compartments, species, reactions and events programatically, and without the need to re-parse the model after each change. Since the re-compile is mandatory for model changes to be realized, we provide an argument called forceRegenerate to all direct API functions which gives users the ability to control when the model is recompiled - i.e. only after all model changes are complete.
\begin{verbatim}
from roadrunner import RoadRunner
rr = RoadRunner(sbmlFile)
rr.addSpecies("A", "cell", 
    initConcentration=5.0, 
    forceRegenerate=False
) 
rr.addReaction("ADeg", ["A"], [], 0.5*A, True)
\end{verbatim}
Example of adding a simple first order mass action degradation reaction to a loaded sbml model. The code assumes the a compartment called {"cell"} was loaded in the initial sbmlFile. Variables: {sbmlFile} is a full path to a valid SBML file on disc.

%
\subsection{Julia Language bindings}
The Julia programming language has gained traction with the systems biology community in recent years and we have therefore implemented language bindings to connect Julia users to libRoadRunner. Whilst our Python bindings are implemented using SWIG \citep{beazley1996swig}, our Julia bindings export symbols from the libRoadRunner shared library which are then imported into Julia using the ccall method below.
\begin{verbatim}
# get julia bindings
import Pkg
Pkg.add("RoadRunner")
# simulate a model
using RoadRunner
rr = RoadRunner.createRRInstance()
RoadRunner.loadSBML(rr, sbmlString)
S = RoadRunner.getFloatingSpeciesIds(rr)
data = RoadRunner.simulateEx(rr, 0, 40, 500)
\end{verbatim}
An illustration showing how to load an SBML model and perform a simulation. The first two lines installs the libRoadRunner language bindings in Julia and the rest of the code compiles an SBML model {sbmlString} and runs a simulation using the {simulateEx} method
%
\subsection{Plugin System}
We have developed a robust plugin system, along with some examples in the form of parameter estimation algorithms that use libRoadRunner solvers and an interface to <which tool?> for bifurcation analysis. 

%
\subsection{Miscellaneous New Functionality}
In previous versions of libRoadRunner, we did not support <input list here>. Now we have built support for everything in the SBML level 3 version 2 specification except for delay differential equations. We have also extended the stoichiometry matrix to …..
Version 2 of libRoadRunner also includes a number of other miscellaneous changes. These include additions to numerical routines used to solve for the time course of differential equations and for computing steady state. In particular we have implemented a basic Euler integration method which has been used for certain time critical applications and an RK45 that can be used to double check the accuracy of the time course solution generated by the default CVODE implementation. 

Like our CVODE implementation, our time series sensitivity implementation uses the popular Sundials package \citep{hindmarsh2005sundials}. Specifically, we have two strategies for solving the sensitivity equations. They can either be solved simultaneously with the system equations \citep{maly1996numerical} or solved using a staggered approach \citep{caracotsios1995sensitivity}.

libRoadRunner version 2 also makes use of the Sundials "kinsol" library for implementing new steady state solvers. Specifically, we use the Inexact Newton approach \citep{brown1987local}. 
%

\section{The "BioModels" Problem}
\begin{figure}[tbp]
    \centering
    \includegraphics[width=0.45\textwidth]{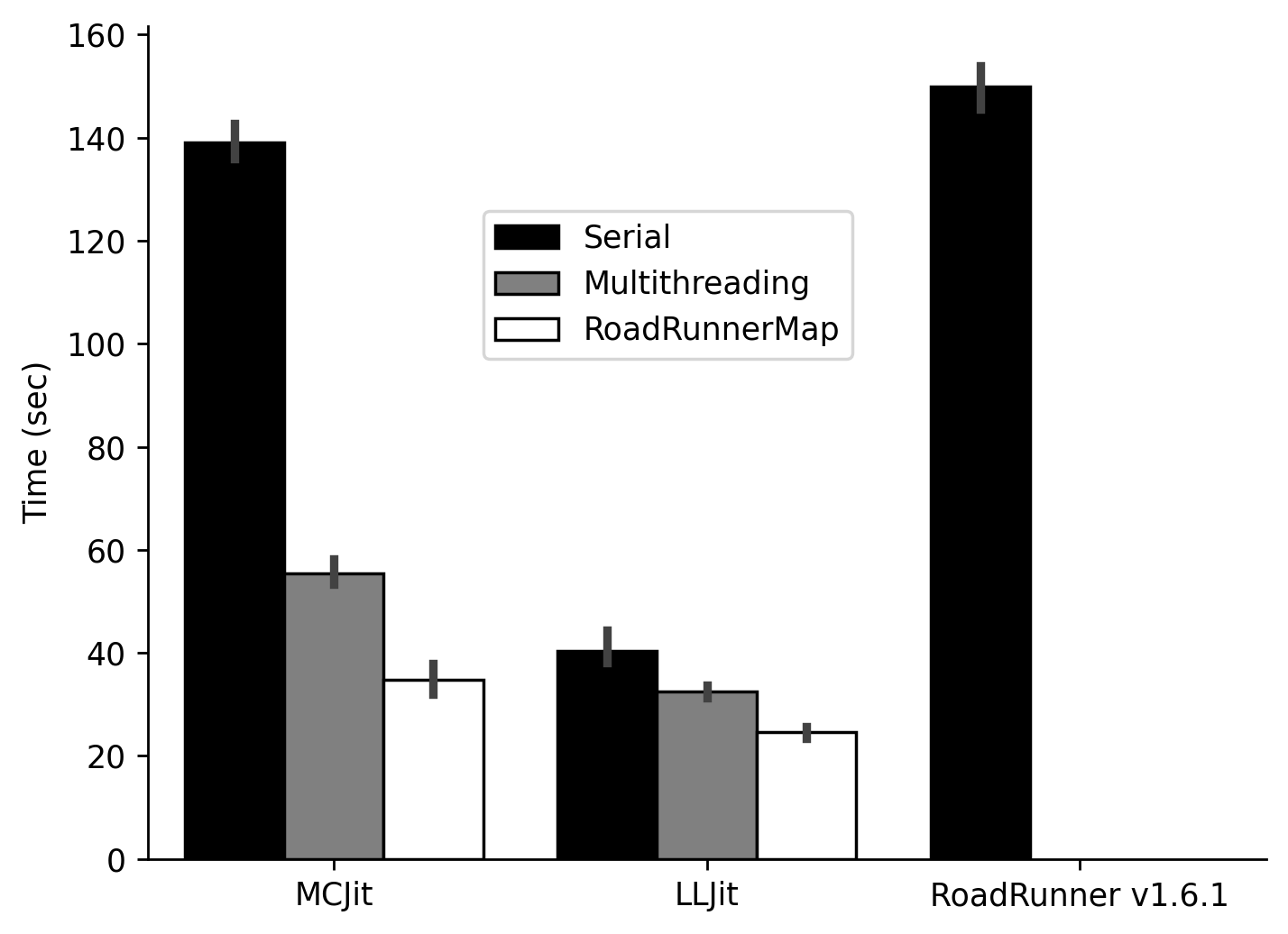}
    \caption{Comparison of time taken for RoadRunner (MCJit or LLJit) or libRoadRunner v1.6.1 to load and simulate a 100 time step time series from the curated section of biomodels (1036 models), either in series or in parallel. Models that failed due to load or simulation errors were ignored (20 for roadrunner, 18 for COPASI). Shown are the means and standard deviations of 5 independent repeats. The number of threads used, where applicable is 12. } 
    \label{fig:biomodels}
\end{figure}

To demonstrate the performance capabilities of libRoadRunner v2, we have measured the time it takes for libRoadRunner v2's MCJit or LLJit compilers and the last implementation of libRoadRunner version 1 (v1.6.1) to load, simulate and store all 1036 models (at the time of writing) from the curated section of the BioModels database \citep{le2006biomodels} in a {RoadRunnerMap}. Moreover, where possible, we repeat this process using {Pool} API from Python's built-in multiprocessing library or using our own {RoadRunnerMap}. We show that libRoadRunner with the {LLJit} compiler is the fastest compiler that we have built to date and that significant performance gains can be had by using the {RoadRunnerMap} for problems involving multiple {RoadRunner} instances \cref{fig:biomodels}.
\section{Discussion}
%
%
libRoadRunner is a fast and convenient tool for both individuals who are investigating the dynamics of a biological system and for tool developers who are building new methods for solving and analysing such systems. In this article we introduce libRoadRunner version 2, where we have built a variety of new tools for the construction, compilation, analysis and solving of dynamical systems described in SBML. 

libRoadRunner version 1 is highly optimized for the simulation of a dynamical system and thanks to our JIT compilation strategy, provides some of the fastest numerical integration routines around \citep{somogyi2015libroadrunner} (other refs for performance metrics?). However, one of the disadvantages of our strategy is that when the need arises for simulation of many SBML models together, run time is dominated by compile time. Examples of such a need include ensemble modelling, where many instances of SBML with varying parameters or topologies need simulating simultaneously. 

To alleviate this bottleneck, in libRoadRunner version 2 we have prioritized new features that enhance the speed with which a model can be compiled. One such feature is an entirely new compiler called {LLJit} which sits side-by-side with the older {MCJit}. We have demonstrated {LLJit} is significantly faster than earlier libRoadRunner implementations (\cref{fig:CompileTime} and \cref{fig:biomodels}) at compiling the same code.

While decreasing compile-time is a worthy goal, there is a natural limit to the speed with which a single model can be compiled. An alternative mechanism for enhancing performance in multi-model problems is to make better use of the available resources that exist in most modern computer systems using parallelism. In libRoadRunner v2, we introduce parallelism in two ways. Firstly we have built a {RoadRunner} container called {RoadRunnerMap} which is capable of orchestrating parallel compiles and secondly we have implemented support for Python's pickle protocol. While the former enables us to abstract parallelism away from the user completely, the latter allows our more computationally competent users to devise their own parallel computation. We have compared these three strategies (serial code, multithreading.Pool and RoadRunnerMap) by loading and simulating a time series from the curated section of BioModels \cref{fig:biomodels}.

Other changes in libRoadRunner version 2 include new steady state solvers, access to time series sensitivities, direct access to model topology 
\section{Conclusion}
In this short article we describe a number of additions to libroadrunner version 2.  These changes are geared towards making libroadrunner more efficient in running, loading and changing models at runtime. These features were added to support a number of specific use cases. These include two main applications: parameter optimization on large compute clusters, and using libRoadRunner to create large model ensembles that includes variation in parameters as well as rate laws and network topology. 

\section{Availability}
libRoadRunner is available for  Windows, Mac OS and Linux operating systems. Precompiled binaries are available from our GitHub webpage \url{https://github.com/sys-bio/roadrunner/releases} and our Python front end is available with {pip install libroadrunner}. The Julia source code is available at \url{https://github.com/sys-bio/RoadRunner.jl} and can be installed via Julia's package manager with {\tt Pkg.add("Road\-Runner")}.
\section{Acknowledgments}

The authors are most grateful to generous funding from the National Institute of General Medical Sciences of the National Institutes of Health under award R01 GM-123032. The content is solely the responsibility of the authors and does not necessarily represent the official views of the National Institutes of Health or the University of Washington. We would like to thank the many users who have contributed to libroadrunner via suggestions for improvements and bug reports. 

\vspace{14pt}
\bibliographystyle{natbib}
\bibliography{document}
\end{document}